\documentclass[a4paper,fleqn]{cas-sc}

\usepackage[authoryear,longnamesfirst]{natbib}

\usepackage{comment}
\usepackage{bm}
\usepackage{siunitx}
\usepackage{chngcntr}

\def\tsc#1{\csdef{#1}{\textsc{\lowercase{#1}}\xspace}}
\tsc{WGM}
\tsc{QE}

\begin{document}
\let\WriteBookmarks\relax
\def\floatpagepagefraction{1}
\def\textpagefraction{.001}

\shorttitle{}

\shortauthors{}

\title [mode = title]{Sensitivity-Guided Parameter Estimation of Bubbly Liquid via a Nonlinear Wave Equation and Physics-Informed Neural Networks}



%

\author[1]{Takeru Hasegawa}[orcid=0000-0003-4064-7178]




\credit{}

\affiliation[1]{organization={Graduate School of Systems and Information Engineering, University of Tsukuba},
            addressline={1-1-1 Tennodai}, 
            city={Tsukuba},
            postcode={305-8573}, 
            country={Japan}}

\author[2]{Naoto Mitsume}[orcid=0000-0002-1090-6573]




\credit{}

\affiliation[2]{organization={Institute of Systems and Information Engineering, University of Tsukuba},
            addressline={1-1-1 Tennodai}, 
            city={Tsukuba},
            postcode={305-8573}, 
            country={Japan}}

\author[2]{Shota Deguchi}[orcid=0000-0002-9538-8663]




\credit{}

\author[3]{Mitsuteru Asai}




\credit{}

\author[2]{Tetsuya Kanagawa}[orcid=0000-0001-8162-2060]





\credit{}




\begin{abstract}
Interactions between ultrasound and microbubbles underlie a range of biomedical and industrial applications, yet predicting how the physical parameters of the medium and bubbles govern nonlinear wave propagation remains difficult. This study proposes a parameter estimation framework that combines a Korteweg--de Vries--Burgers (KdVB) equation, describing weakly nonlinear wave propagation in a bubbly liquid, with a physics-informed neural network (PINN). The contributions are threefold. First, the PINN reproduces the reference solution of the governing equation in the forward problem, with a near-zero prediction error at the initial time that confirms the initial condition is enforced exactly as intended. Second, prior to inverse estimation, a sensitivity analysis is introduced that quantifies, at both the coefficient and solution levels, how each of eight physical parameters influences the KdVB equation, allowing the expected difficulty of estimating each parameter to be assessed in advance. Third, based on this assessment, single-parameter inverse estimation is performed for seven of the eight parameters using the PINN. Liquid viscosity, bubble initial radius, liquid density, sound speed, and liquid pressure are estimated with errors of 0.8--2.1\%, the polytropic exponent converges only partially, and the surface tension coefficient cannot be estimated at all, consistent with the sensitivity analysis. These results establish a general framework for identifying and estimating the physical parameters governing nonlinear wave propagation in bubbly liquids from observed waveform data.
\end{abstract}


\begin{highlights}
\item PINN combined with a KdVB equation estimates physical parameters of a bubbly liquid.
\item PINN accurately reproduces the KdVB reference solution in the forward problem.
\item Coefficient- and solution-level sensitivity analysis predicts parameter estimability.
\item Five of seven physical parameters are estimated with errors of 0.8--2.1\%.
\end{highlights}


\maketitle

\section{Introduction} \label{sec:intro}
Interactions between ultrasound and bubbles have been widely investigated and applied across a variety of fields, including ultrasonic cleaning \citep{MASON2016519,VERHAAGEN2016619,REUTER2017542}, food processing \citep{BHARGAVA2021105293,SINGLA2021105506}, the enhancement of chemical reactions via sonochemical effects \citep{Kenneth1990,Savun-Hekimoglu2040042}, and biomedical engineering \citep{Ohl2015,CHOWDHURY202075}. In particular, in the field of biomedical engineering, growing attention has been paid to applications that combine micro- and nanoscale bubbles with ultrasound, such as drug delivery systems \citep{CHOWDHURY202075,Fan2014}, the generation of ultrasound contrast agents \citep{CHATTERJEE20031749,GUMMER20212958}, and focused ultrasound therapies, including lithotripsy and tumor ablation \citep{Baker2001,terHaar1999,Kennedy2003}.

Focused ultrasound therapy is a minimally invasive treatment modality compared with conventional surgical resection and radiation therapy, and has been applied to tumor ablation and lithotripsy \citep{Baker2001,terHaar1999,Kennedy2003,Fry1995}. In recent years, combining microbubbles with focused ultrasound has been reported to enhance the therapeutic effect \citep{CHUNG2012e519}; however, excessive irradiation dose or power, or an inappropriate amount of introduced bubbles, may damage healthy tissue surrounding the target region \citep{Hasan2021,Sherwin2003,Sturtevant1998}. If the behavior of ultrasound propagating through the target medium could be accurately predicted in advance and irradiation conditions designed accordingly, such risks could be mitigated. However, empirical observation alone is insufficient to systematically understand and control the effects of physical parameters, such as the material properties of the medium and the initial state of the bubbles, on ultrasound propagation. Furthermore, parametrically exploring this dependence through numerical simulation can incur substantial computational cost. A technique capable of directly estimating the governing physical parameters from observed wave information, without relying on numerous numerical simulations, is therefore expected to considerably contribute to a systematic understanding and control of ultrasound propagation characteristics, and, in turn, to the design of safe irradiation conditions.

A key element enabling such inverse estimation is a theoretical model that describes ultrasound propagation in a medium containing bubbles. Reduced-order nonlinear wave equations describing one-dimensional nonlinear ultrasound propagation in a medium containing bubbles have been derived using singular perturbation methods, for media ranging from viscoelastic materials to liquids \citep{Wijngaarden_1968,Wijngaarden1972,Kanagawa2010,Hasegawa2023,Kurata2025,yu2025propagation}. A distinguishing feature of these models is that the coefficients governing the nonlinear, dissipative, and dispersive effects are expressed explicitly as functions of physical parameters, such as the material properties of the medium and the initial state of the bubbles. The Korteweg--de Vries--Burgers (KdVB) equation employed in this study \citep{yu2025propagation} belongs to this theoretical lineage and provides a framework for describing wave propagation in a bubbly liquid. More broadly, the KdVB equation has been used to model a range of physical phenomena beyond bubbly liquids, such as undular bores in shallow water, wave propagation in viscous fluid-filled elastic tubes, and weakly nonlinear plasma waves with dissipative effects, among others \citep{feng2007burgers}. By enabling a quantitative evaluation of how each physical parameter contributes to the wave propagation characteristics, this framework provides the theoretical basis for an inverse-problem-based approach in which physical parameters are estimated from observed waveforms.

Physics-informed neural networks (PINNs) \citep{RAISSI2019686} have recently attracted attention as a powerful approach to such inverse problems. By incorporating physical information, such as governing equations, initial conditions, and boundary conditions, into the loss function, PINNs enable both forward problems (predicting the solution of the governing equations) and inverse problems (estimating parameters within the equations) to be addressed within a unified, automatic-differentiation-based framework, at a computational cost comparable to that of the forward problem alone. PINNs have accordingly been applied to weakly nonlinear wave equations of this class: Bhaumik et al. \citep{bhaumik2024effects} used a semi-supervised PINN to solve the forward problem for perturbed Burgers, Korteweg--de Vries (KdV), and KdVB equations, derived via a reductive perturbation approach for weakly nonlinear wave transmission in a viscoelastic tube; more closely related to the present study, Kang et al. \citep{kang2025physics} applied a PINN to a KdV model of internal solitary waves in a two-layer fluid, both learning a parameter-to-solution operator and estimating the physical parameters embedded in its coefficients, namely the layer-depth ratio and the fluid densities. However, in contrast to the dispersive-only KdV equation considered by Kang et al. \citep{kang2025physics}, the KdVB equation adopted in this study additionally incorporates a dissipative term. Identifiability and sensitivity assessments have also been combined with PINN-based parameter estimation in other contexts, such as ordinary-differential-equation models in systems biology \citep{daneker2023systems}, but, to the best of the authors' knowledge, such an assessment has not been performed, prior to inverse estimation, for a PINN applied to a weakly nonlinear wave equation of the kind considered here.

In this study, we therefore construct a parameter estimation framework that combines the KdVB equation, which describes nonlinear ultrasound propagation in a bubbly liquid, with a PINN. The contributions of this study are threefold. First, we show that the PINN can accurately predict the solution of the governing equation (forward problem). Second, prior to performing inverse estimation, we present a sensitivity analysis framework, in which the sensitivity of both the governing-equation coefficients and the solution itself to each physical parameter is evaluated in order to quantitatively assess, in advance, the expected difficulty of estimating each parameter. Third, building on this prior assessment, we perform single-parameter inverse estimation and confirm that the resulting estimation accuracy is consistent with the sensitivity analysis.

The remainder of this paper is organized as follows. Section~\ref{sec:govern_eq} introduces the KdVB equation and its relation to the eight governing physical parameters. Section~\ref{sec:methodology} presents the PINN framework that incorporates the KdVB equation as physical information. Section~\ref{sec:results_discussion} validates the proposed approach through the forward problem, the sensitivity analysis, and the inverse problem introduced above. Section~\ref{sec:summary} summarizes the findings of this study and discusses future perspectives.

\section{Governing equation} \label{sec:govern_eq}
We use the KdVB equation for nonlinear wave propagation in a bubbly liquid \citep{yu2025propagation}.
\begin{align}
&\frac{\partial u}{\partial t} 
+ \Pi_1 u\frac{\partial u}{\partial x} 
- \Pi_2 \frac{1}{\sqrt{\epsilon}v_\mathrm{p}} \frac{\partial^2 u}{\partial x^2}
- \Pi_3 \frac{1}{\epsilon v^2_\mathrm{p}} \frac{\partial^3 u}{\partial x^3}
= 0, \quad
x \in [-1, 1], \quad
t \in [0, 1], \nonumber \\
&\mathrm{Initial~condition:~} u(x,0) = u_0(x) = 0.1 \sin{(\pi x)}, \nonumber \\ 
&\mathrm{Periodic~boundary~condition~at~} x = \pm 1,
\label{eq:kdvb}
\end{align}
where \(t\) is time, \(x\) is the space coordinate, \(0 < \epsilon \ll 1\) is a dimensionless parameter that defines the strength of nonlinearity, and \(v_\mathrm{p}\) is the nondimensional phase velocity. \(\Pi_1\), \(\Pi_2\), and \(\Pi_3\) are the coefficients of nonlinear, dissipative, and dispersive terms, respectively, the balance of which governs the characteristics of nonlinear wave propagation. These coefficients are governed by eight physical constant parameters: the bubble initial radius \(R_0^\ast\), the bubble number density \(N_\mathrm{g}^\ast\), the polytropic exponent of gas \(\gamma\), the density of liquid without disturbance \(\rho_\mathrm{l0}^\ast\), the sound speed in pure water \(c_\mathrm{l0}^\ast\), the surface tension coefficient \(\sigma^\ast\), the viscosity of liquid \(\mu_\mathrm{l}^\ast\), and the pressure of liquid without disturbance \(p_\mathrm{l0}^\ast\). These coefficients can be written explicitly in terms of these eight parameters alone (see \ref{sec:explicit_form}), and are therefore independent of the space-time coordinates \(x\) and \(t\). Here, the superscript \(\ast\) indicates a dimensional quantity; the absence of this superscript indicates a nondimensional quantity.

\begin{figure}[t]
  \centering
    \includegraphics[width=\textwidth]{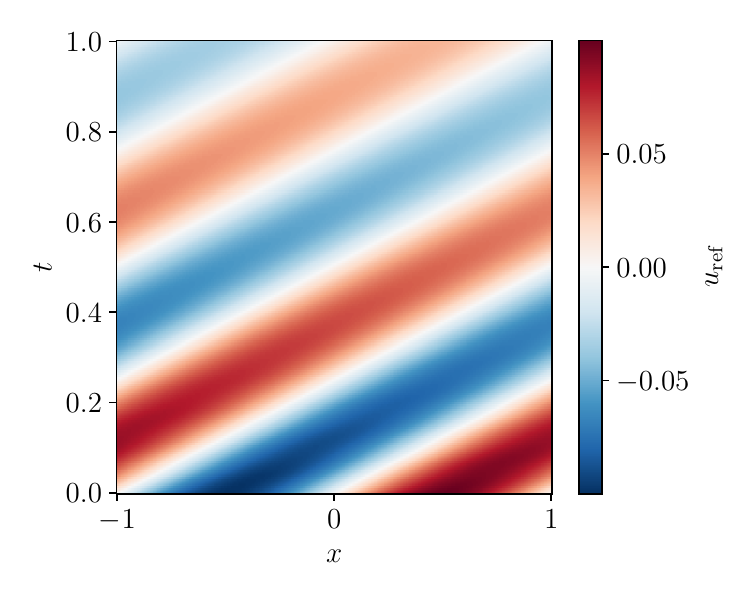}
    \caption{Reference solution of the governing equation (\ref{eq:kdvb}).}\label{fig:fwd_m2_reference}
\end{figure}

\section{Methodology} \label{sec:methodology}
In a standard PINN framework \citep{RAISSI2019686}, the neural network directly outputs an approximation \(\tilde{u}(x,t) = \mathrm{NN}(x,t;\bm{\theta})\) of the solution, where \(\bm{\theta}\) denotes trainable parameters in the neural network, and the initial and boundary conditions are enforced only as additional penalty terms in the loss function,
\begin{align}
\mathcal{L}
= \mathcal{L}_\mathrm{pde} + \mathcal{L}_\mathrm{data} + \mathcal{L}_\mathrm{ic} + \mathcal{L}_\mathrm{bc},
\label{eq:standard_pinn_loss}
\end{align}
where
\begin{align}
\mathcal{L}_\mathrm{ic}
=
\frac{1}{N_\mathrm{ic}}\sum^{N_\mathrm{ic}}_{i=1}\left|\tilde{u}\left(x^i_\mathrm{ic}, 0\right) - u_0\left(x^i_\mathrm{ic}\right)\right|^2
\label{eq:standard_pinn_ic}
\end{align}
penalizes deviation from the initial condition, and
\begin{align}
\mathcal{L}_\mathrm{bc}
=
\frac{1}{N_\mathrm{bc}}\sum^{N_\mathrm{bc}}_{i=1}
\sum^{K}_{k=0}
\left|
\frac{\partial^k \tilde{u}\left(1, t^i_\mathrm{bc}\right)}{\partial x^k}
-
\frac{\partial^k \tilde{u}\left(-1, t^i_\mathrm{bc}\right)}{\partial x^k}
\right|^2
\label{eq:standard_pinn_bc}
\end{align}
penalizes deviation from periodicity of \(\tilde{u}\) and of its spatial derivatives up to the order \(K\) required by the governing equation, evaluated at the domain boundaries \(x = \pm 1\). Because these conditions are enforced only as soft penalties, minimizing \(\mathcal{L}\) does not guarantee that they are satisfied exactly, so residual violations of the initial and periodic boundary conditions can persist even after training.

To avoid this issue, in this study the periodic boundary condition and the initial condition of the governing equation (\ref{eq:kdvb}) are instead enforced exactly, at the architectural level, rather than penalized as in Eqs.~(\ref{eq:standard_pinn_ic}) and (\ref{eq:standard_pinn_bc}). For the periodic boundary condition, the Fourier feature embedding method is used \citep{dong2021exactlyperiodic}, 
\begin{align}
v(x) 
= 
\left[
1, 
\sin{\left(\pi x\right)}, \cos{\left(\pi x\right)}, 
\sin{\left(2\pi x\right)}, \cos{\left(2\pi x\right)},
...,
\sin{\left(m\pi x\right)}, \cos{\left(m\pi x\right)} 
\right]^\mathrm{T} \in \mathbb{R}^{2m+1}.
\label{eq:fourier_feature_embeddings}
\end{align}
Since \(v(x)\) is exactly periodic in \(x\), any network built on this embedding is exactly periodic. For the initial condition, the output of the neural network is multiplied by time and added to the initial condition function as
\begin{align}
\hat{u}(x,t) = u_0(x) + t \cdot \mathrm{NN}(v(x), t; \bm{\theta}),
\label{eq:pinn_arch}
\end{align}
where the neural network (NN) architecture takes \(\left(2m+2\right)\)-dimensional input and provides a one-dimensional output composing \(\hat{u}\). Detailed specifications of the network architecture are provided in Section \ref{sec:results_discussion}. Because \(\hat{u}(x,0) = u_0(x)\) identically and \(\hat{u}\) is exactly periodic in \(x\) by construction, both the initial condition and the periodic boundary condition are satisfied exactly and the penalty terms \(\mathcal{L}_\mathrm{ic}\) and \(\mathcal{L}_\mathrm{bc}\) in Eq.~(\ref{eq:standard_pinn_loss}) are no longer required. The loss function \(\mathcal{L}\) therefore reduces to only the residual term of the governing equation \(\mathcal{L}_\mathrm{pde}\) and the data-fitting error term \(\mathcal{L}_\mathrm{data}\) as
\begin{align}
\mathcal{L} 
&= \mathcal{L}_\mathrm{pde} + \mathcal{L}_\mathrm{data} \nonumber \\
&= 
\frac{1}{N_\mathrm{pde}}\sum^{N_\mathrm{pde}}_{i=1}\left|f\left(\hat{u}\left(x^i_\mathrm
{pde}, t^i_\mathrm{pde}\right)\right)\right|^2
+
\frac{1}{N_\mathrm{data}}\sum^{N_\mathrm{data}}_{j=1}\left|\hat{u}\left(x^j_\mathrm{data}, t^j_\mathrm{data}\right) - u^j_\mathrm{data}\right|^2
,
\end{align}
where \(f(u)\) is the residual of the governing equation (\ref{eq:kdvb}),
\begin{align}
f(u) 
= 
\frac{\partial u}{\partial t} 
+ \Pi_1 u\frac{\partial u}{\partial x} 
- \Pi_2 \frac{1}{\sqrt{\epsilon}v_\mathrm{p}} \frac{\partial^2 u}{\partial x^2}
- \Pi_3 \frac{1}{\epsilon v^2_\mathrm{p}} \frac{\partial^3 u}{\partial x^3}.
\end{align}
\(\{(x^i_\mathrm{pde}, t^i_\mathrm{pde})\}^{N_\mathrm{pde}}_{i=1}\) and 
\(\{((x^j_\mathrm{data}, t^j_\mathrm{data}), u^j_\mathrm{data})\}^{N_\mathrm{data}}_{j=1}\) are the sets of collocation points and observational data, respectively. 

\begin{figure}[t]
  \centering
    \includegraphics[width=\textwidth]{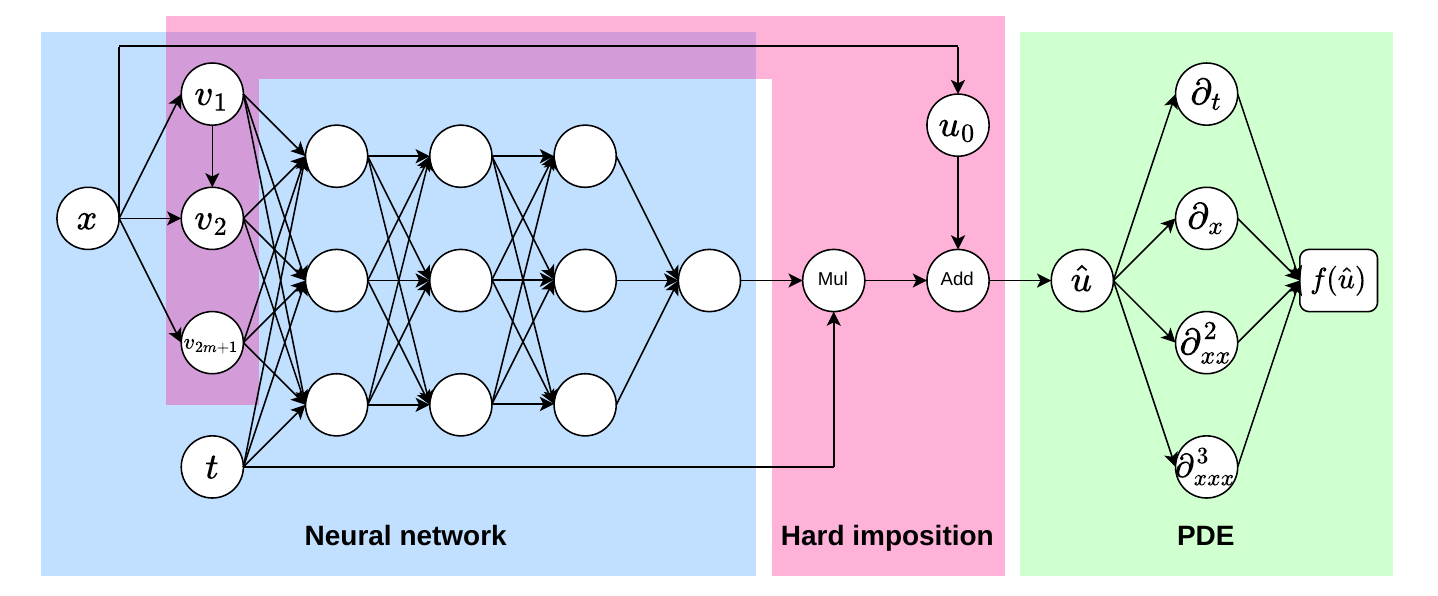}
    \caption{Schematic diagram of the architecture of PINN developed in this study.}\label{fig:pinn_arch}
\end{figure}

\section{Results and discussion} \label{sec:results_discussion}
We evaluated whether the PINN developed in this study can appropriately learn the weakly nonlinear wave propagation phenomena in bubbly liquids, and whether it can estimate key parameters governing the phenomenon, such as the physical properties of the medium and the initial bubble radius. To properly interpret the estimation results, we first performed a sensitivity analysis to assess, in advance, the expected difficulty of estimating each parameter.

Our PINN model is developed using the PyTorch framework \citep{paszke2019pytorch}, and training is conducted on an NVIDIA GeForce RTX 4070. The degree of Fourier feature embeddings is \(m = 2\). The neural network (NN) in Eq.~(\ref{eq:pinn_arch}) then passes the input feature through six fully connected hidden layers, each with 64 neurons, and provides a scalar output. Hence, the architecture is \([2m+2, 64, 64, 64, 64, 64, 64, 1]\). The tanh function is used as the activation function in hidden layers. The numbers of collocation points and observational data are \(N_\mathrm{pde} = 3{,}000\) and \(N_\mathrm{data} = 600\), respectively; the spatiotemporal coordinates of both are randomly resampled at each epoch, and the corresponding values of the observational data are obtained by interpolating the reference solution shown in Fig.~\ref{fig:fwd_m2_reference}. The Adam optimizer is used for training. The learning rate is initially set to \(0.001\), and halved every \(20{,}000\) epochs. This is to maintain a relatively large learning rate during the early stages of the training, while gradually reducing it to carefully tune the model in the later stages.

\begin{table}[t]
\caption{Physical constant parameters and theoretical parameters used in the numerical experiments. The values are taken from Ref.~\citep{yu2025propagation}.}
\label{tbl:physical_constant_theoretical_parameter}
\begin{tabular*}{\tblwidth}{@{}LLL@{}}
\toprule
 Parameter & Symbol & Value \\ 
\midrule
\textbf{Physical constant} \\
Bubble initial radius & \(R^\ast_0\) & \qty{4.5}{\micro\meter} \\
Bubble number density & \(N^\ast_\mathrm{g}\) & \qty{2e11}{\meter^{-3}} \\
Polytropic exponent of gas & \(\gamma\) & \qty{1.4}{} \\
Density of liquid without disturbance & \(\rho^\ast_\mathrm{l0}\) & \qty{997}{\kilogram/\meter^3} \\
Sound speed in pure water & \(c^\ast_\mathrm{l0}\) & \qty{1481}{\meter/\second} \\
Surface tension coefficient & \(\sigma^\ast\) & \qty{0.0725}{\newton/\meter} \\
Viscosity of liquid & \(\mu^\ast_\mathrm{l}\) & \qty{0.001}{\pascal\second} \\
Pressure of liquid without disturbance & \(p^\ast_\mathrm{l0}\) & \qty{1.01325e5}{\pascal} \\
\textbf{Theoretical parameter} \\
Parameter related to strength of nonlinearity & \(\epsilon\) & \qty{0.05}{} \\
Parameter related to wave frequency & \(\omega\) & \qty{0.2}{} \\
\bottomrule
\end{tabular*}
\end{table}

\subsection{Forward problem}
\begin{figure}[t]
  \centering
    \includegraphics[width=\textwidth]{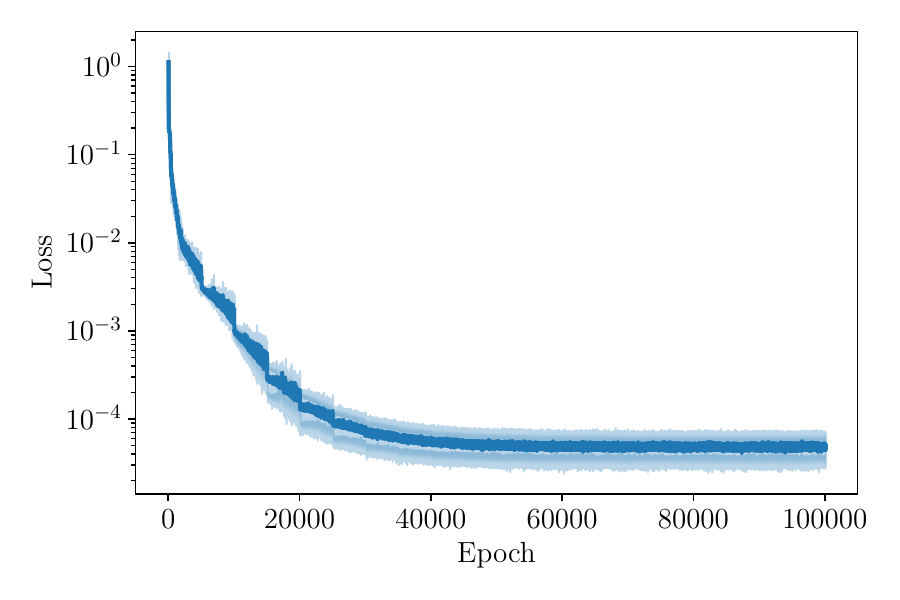}
    \caption{History of the total loss function in the forward problem. The solid line represents the mean value over five runs with different random seeds, and the shaded region represents the corresponding standard deviation.}\label{fig:fwd_m2_loss}
\end{figure}

For the forward problem, we trained a neural network, which incorporates the KdVB equation (\ref{eq:kdvb}) and the parameter settings listed in Table \ref{tbl:physical_constant_theoretical_parameter} as physical information, as a solver for 100,000 epochs, and evaluated its training performance. To assess the influence of the neural network's dependence on initial parameter values, training was performed using five different random seeds. Figure \ref{fig:fwd_m2_loss} shows the evolution of the total loss \(\mathcal{L}\). The loss decreases rapidly during the early stages of training, then gradually levels off, becoming nearly flat after 50,000 epochs. This indicates that 100,000 epochs were sufficient for training the neural network.

\begin{figure}[t]
  \centering
  \begin{minipage}[t]{0.48\columnwidth}
    \includegraphics[width=\columnwidth]{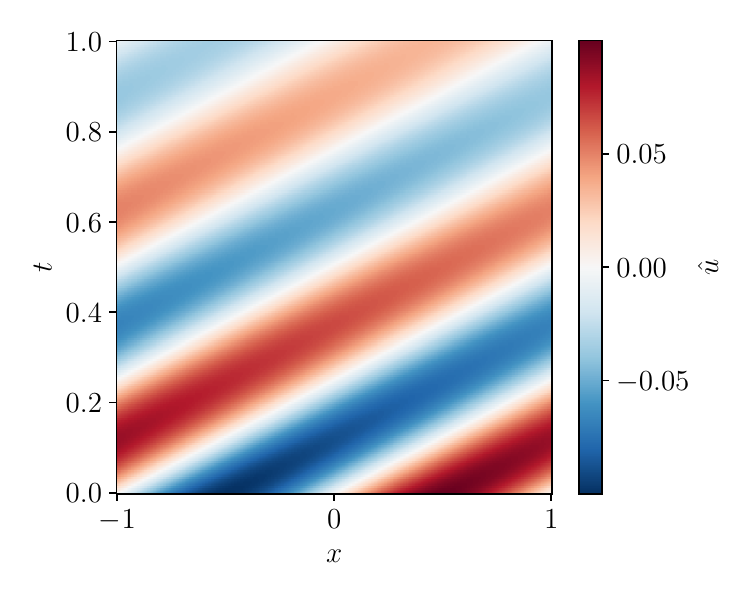}
  \end{minipage}
  \begin{minipage}[t]{0.48\columnwidth}
    \includegraphics[width=\columnwidth]{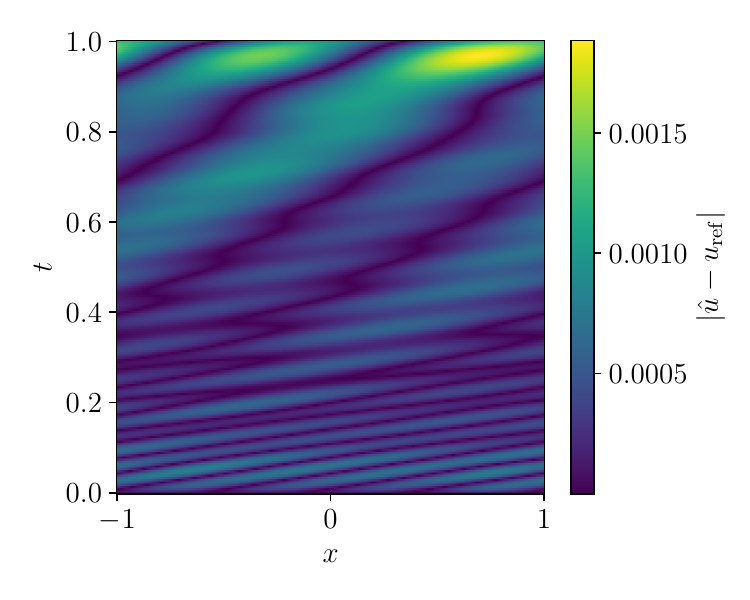}
  \end{minipage}
    \caption{PINN prediction of the solution of Eq.~(\ref{eq:kdvb}), \(\hat{u}\), and absolute error between the PINN prediction and the reference solution \(\left|\hat{u} - u_\mathrm{ref}\right|\). The PINN prediction is obtained as the mean of the predictions over five runs with different random seeds.}\label{fig:fwd_m2_spacetime}
\end{figure}

Figure \ref{fig:fwd_m2_spacetime} shows the absolute error between the PINN prediction after 100,000 epochs of training and the reference solution shown in Fig. \ref{fig:fwd_m2_reference}. The PINN prediction can be seen to reproduce the overall trend of the reference solution. On the other hand, the absolute error is observed to increase as time progresses. This behavior commonly arises when PINNs are applied to problems such as advection equations, in which physical quantities propagate across the computational domain over time, and similar phenomena have been reported in previous studies \citep{krishnapriyan2021characterizingpossiblefailuremodes, WANG2022110768,wang2023expertsguidetrainingphysicsinformed}. This issue could potentially be mitigated by introducing techniques such as curriculum learning \citep{krishnapriyan2021characterizingpossiblefailuremodes,wang2023expertsguidetrainingphysicsinformed}.

\begin{figure}[t]
  \centering
    \includegraphics[width=\textwidth]{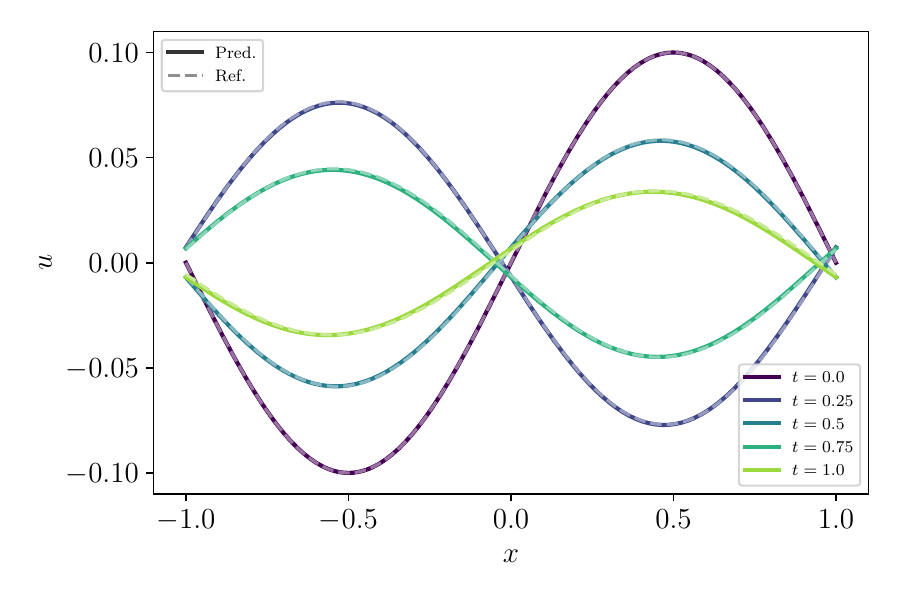}
    \caption{Snapshots of the PINN prediction and the reference solution. The PINN prediction is obtained as the mean of the predictions over five runs with different random seeds.}\label{fig:fwd_m2_snapshots}
\end{figure}

Figure \ref{fig:fwd_m2_snapshots} shows snapshots of the PINN prediction and the reference solution at specific times. The sinusoidal wave imposed as the initial condition propagates and gradually decays over time; at \(t=0.75\) and \(t=1.0\), a clear deviation of the predicted solution from the reference solution can be observed. On the other hand, the nonlinear effect (wave steepening) and the dispersive effect, both of which act on nonlinear waves propagating through a bubbly medium, are found to be almost negligible. Whether these effects become dominant depends not only on the parameter settings listed in Table \ref{tbl:physical_constant_theoretical_parameter} but also on several other factors, including the range of the spatiotemporal coordinates considered and the choice of initial conditions; under the settings adopted in this study, these effects are considered not to be dominant.

\begin{figure}[t]
  \centering
    \includegraphics[width=\textwidth]{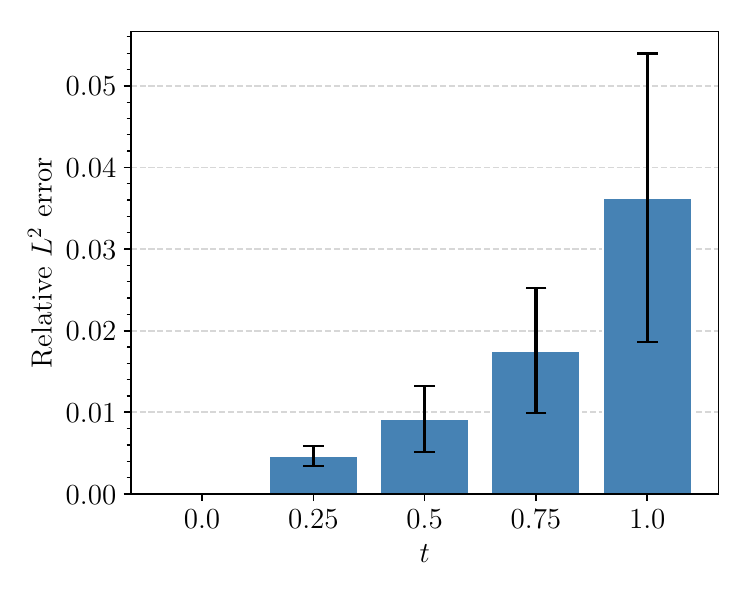}
    \caption{Spatial relative $L^2$ error between the PINN prediction and the reference solution. Bars represent the mean value over five runs with different random seeds, and error bars represent the magnitude of the standard deviation.}\label{fig:fwd_m2_l2bar}
\end{figure}

Figure \ref{fig:fwd_m2_l2bar} shows the relative \(L^2\) error at specific times. As discussed above, the error can be readily seen to increase with time; notably, however, the error at the initial time \(t=0\), including its standard deviation, is extremely small. This suggests that the PINN architecture, designed to strictly enforce the initial condition as shown in Eq.~(\ref{eq:pinn_arch}), is functioning as intended.

\subsection{Sensitivity analysis}
In the sensitivity analysis, we evaluated the extent to which the physical constant parameters listed in Table \ref{tbl:physical_constant_theoretical_parameter} affect the nonlinear, dissipative, and dispersive coefficients, as well as the solution. The sensitivity of each coefficient is defined by 
\begin{align}
S_\mathrm{coef}^{ij} = \frac{\partial \Pi_i}{\partial \phi_j} \cdot \left.\frac{\phi_j}{\Pi_i}\right|_{\bm{\phi}=\bm{\phi}_\mathrm{true}},
\label{eq:sens_coef}
\end{align}
where \(\bm{\phi} = \left[\mu_\mathrm{l}, R_0, \rho_\mathrm{l0}, c_\mathrm{l0}, \gamma, p_\mathrm{l0}, \sigma, N_\mathrm{g}\right]^\mathrm{T}\) and the sensitivity of the solution is defined by 
\begin{align}
 &S^i_\mathrm{sol} = \sqrt{\langle {s^i(x,t)}^2 \rangle_{x,t}}, \\
 &s^i(x,t) = \frac{\partial u}{\partial \phi_i} \cdot \left.\frac{\phi_i}{\mathrm{max}_{x,t}\left|u(x,t)\right|}\right|_{\bm{\phi}=\bm{\phi}_\mathrm{true}}. \label{eq:sens_sol}
\end{align}
Here, all physical constants are normalized by their respective true values.

The partial derivatives in Eqs.~(\ref{eq:sens_coef}) and (\ref{eq:sens_sol}) are evaluated numerically using the central difference scheme with a step size of \(h = 10^{-4}\) relative to the normalized parameter,
\begin{align}
\left.\frac{\partial \Pi_i}{\partial \phi_j}\right|_{\bm{\phi}=\bm{\phi}_\mathrm{true}}
\approx
\frac{\Pi_i(\phi_j(1+h)) - \Pi_i(\phi_j(1-h))}{2h\phi_j},
\label{eq:central_diff}
\end{align}
and analogously for \(\partial u/\partial \phi_i\), where \(u(x,t;\phi_i)\) is obtained by solving the governing equation (\ref{eq:kdvb}) with the reference numerical solver at the perturbed parameter value. Although \(\Pi_i\) admit closed-form expressions (\ref{sec:explicit_form}) and could in principle be differentiated analytically, the same central difference scheme is applied to both \(\Pi_i\) and \(u\) for methodological consistency, since \(u\) has no closed-form dependence on \(\bm{\phi}\).

\begin{figure}[t]
  \centering
    \includegraphics[width=\textwidth]{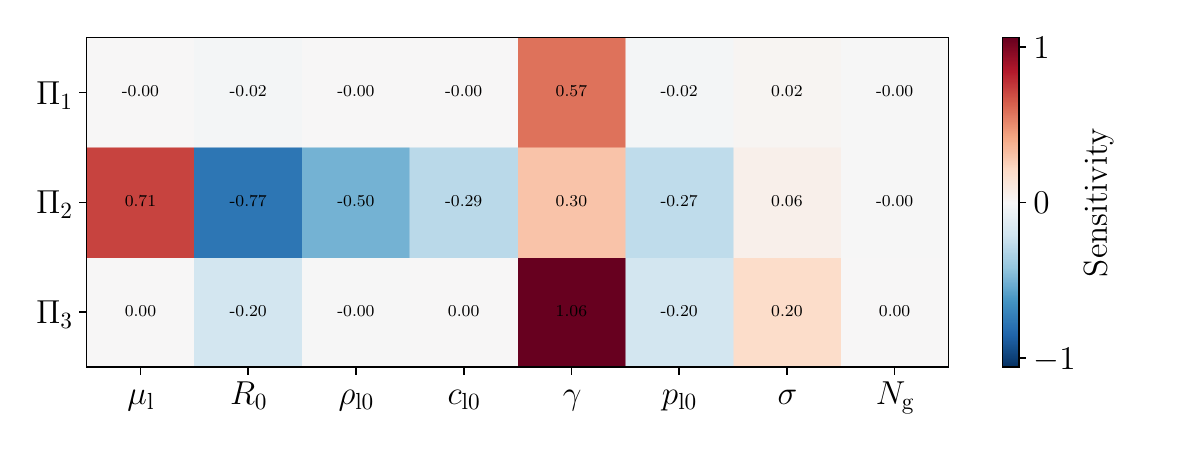}
    \caption{Sensitivity of the coefficients to the physical constants listed in Table \ref{tbl:physical_constant_theoretical_parameter}.}\label{fig:sensitivity_heatmap}
\end{figure}

Figure \ref{fig:sensitivity_heatmap} shows the sensitivity of each coefficient, representing the percentage change in each coefficient resulting from a 1\% change in a given physical constant parameter from its true value listed in Table \ref{tbl:physical_constant_theoretical_parameter}. Examining the results by coefficient, most parameters are found to affect the dissipative and dispersive coefficients, whereas the nonlinear coefficient shows little to no dependence on any parameter other than the polytropic exponent \(\gamma\). In the context of parameter estimation, this implies that the nonlinear coefficient contributes almost nothing to the estimation, which is instead governed primarily by the dissipative and dispersive coefficients. Examining the results by parameter, on the other hand, most parameters are found to affect the dissipative and dispersive coefficients; notably, however, the bubble number density \(N_\mathrm{g}\) exhibits no sensitivity at all. This is not merely a case of extremely small sensitivity but a genuine absence of sensitivity: as shown in  \ref{sec:independent_of_ng}, none of the coefficients in the governing equation depend on \(N_\mathrm{g}\).

\begin{figure}[t]
  \centering
    \includegraphics[width=\textwidth]{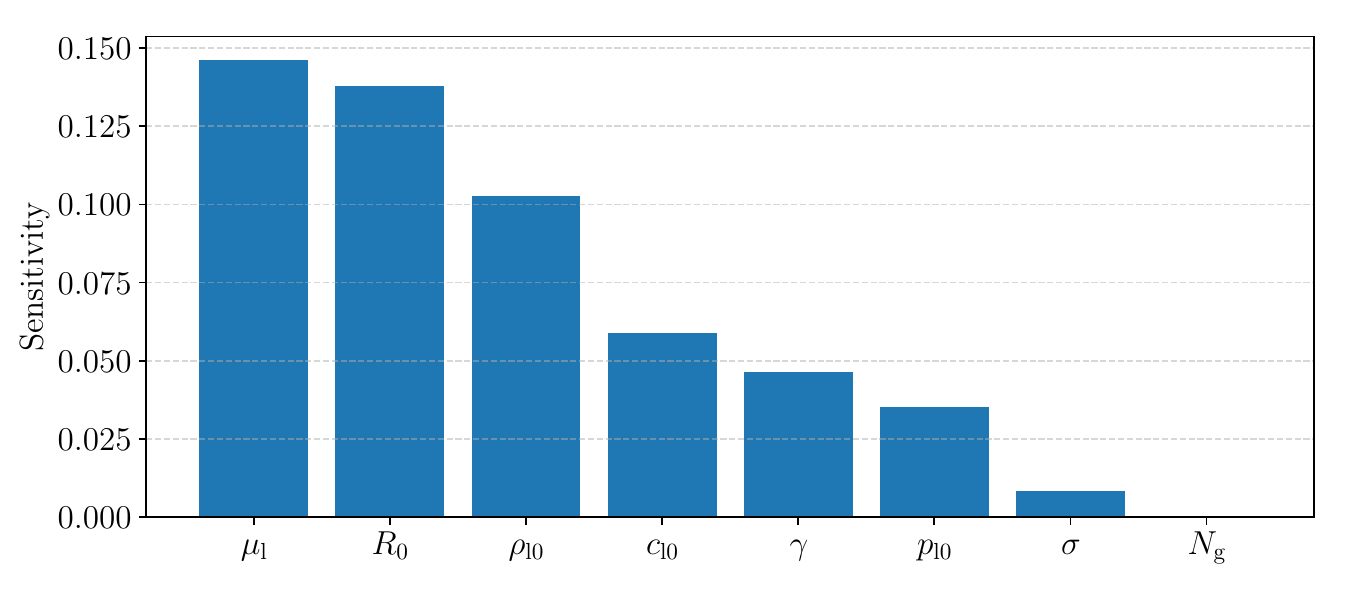}
    \caption{Sensitivity of the solution to the physical constants listed in Table \ref{tbl:physical_constant_theoretical_parameter}.}\label{fig:sensitivity_solution_bar}
\end{figure}

We next discuss the sensitivity of the solution. Since the loss function is the quantity being optimized in our neural network, and since it is defined in terms of the predicted solution and the residuals computed from it, the sensitivity of the solution can be regarded as more fundamentally related to parameter estimation than the sensitivity of the individual coefficients. Figure \ref{fig:sensitivity_solution_bar} shows the sensitivity of the solution, representing the percentage change in the solution resulting from a 1\% change in a given physical constant parameter from its true value listed in Table \ref{tbl:physical_constant_theoretical_parameter}. The magnitude of the sensitivity to each parameter shown in this figure tends to correlate with the difficulty of estimating that parameter. That is, we may expect that the viscosity of liquid \(\mu_\mathrm{l}\) and the bubble initial radius \(R_0\) are likely to be estimated successfully, whereas the surface tension coefficient \(\sigma\) is likely to be more difficult to estimate than the other parameters. As for the bubble number density, Fig. \ref{fig:sensitivity_solution_bar} likewise shows no sensitivity at all; in fact, the governing equation, together with the initial and boundary conditions, is entirely independent of \(N_\mathrm{g}\) (see \ref{sec:independent_of_ng}). Estimating the bubble number density is therefore fundamentally impossible in principle, and it is accordingly excluded from the set of parameters to be estimated in the inverse problem.

\subsection{Inverse problem}
In the inverse problem, we attempted single-parameter estimation for each of the seven physical constant parameters listed in Table \ref{tbl:physical_constant_theoretical_parameter}, excluding the bubble number density. The parameter to be estimated was normalized by its true value, and estimation was performed starting from an initial value of either 2.0 or 0.5 times the true value; that is, an estimated value closer to 1.0 indicates higher estimation accuracy. To ensure that the estimated parameter remains physically valid (i.e., strictly positive) throughout training, the normalized parameter is not optimized directly; instead, a scalar \(\eta\) is introduced as the trainable variable, and the normalized parameter is recovered via the exponential map \(\hat\phi/\phi_\mathrm{true} = \exp(\eta)\), which guarantees \(\hat\phi/\phi_\mathrm{true} > 0\) for any real \(\eta\). The trainable variable is initialized as \(\eta_0 = \ln(2.0)\) or \(\ln(0.5)\), corresponding to the two initial values considered in this study. All parameters other than the one being estimated were fixed at their true values. As in the forward problem, training and estimation were performed for 100,000 epochs using five different random seeds to assess the influence of the neural network's dependence on initial parameter values.

\begin{figure}[t]
  \centering
  \begin{minipage}[t]{0.48\columnwidth}
    \includegraphics[width=\columnwidth]{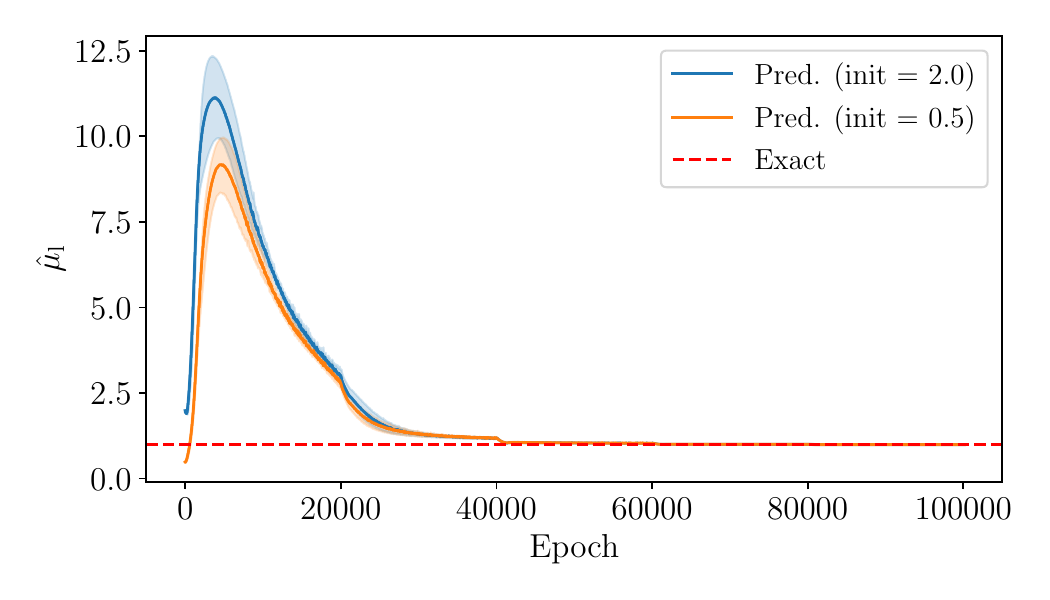}
  \end{minipage}
  \begin{minipage}[t]{0.48\columnwidth}
    \includegraphics[width=\columnwidth]{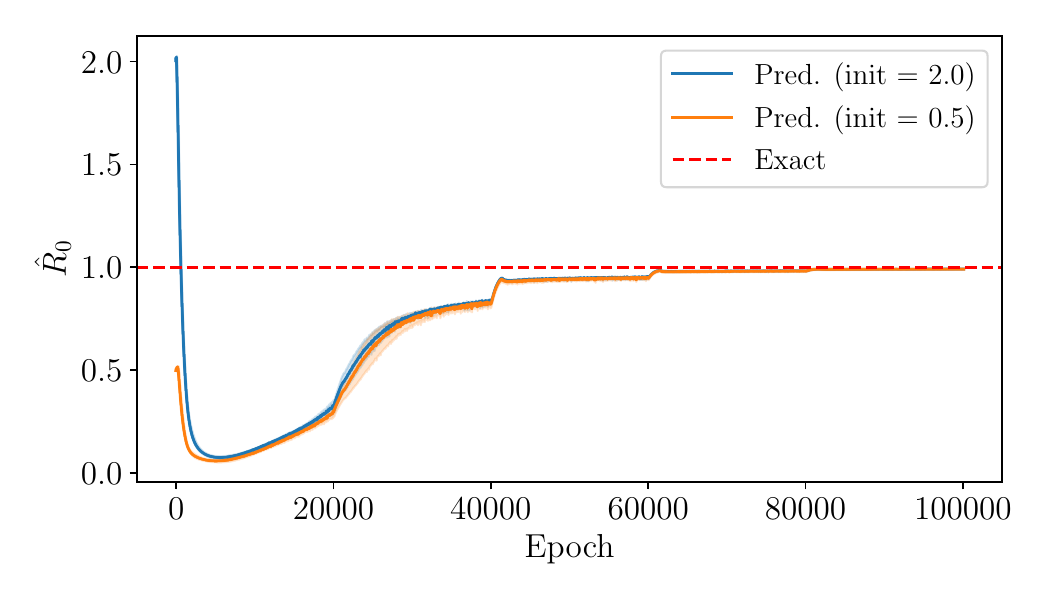}
  \end{minipage}
  \\
  \begin{minipage}[t]{0.48\columnwidth}
    \includegraphics[width=\columnwidth]{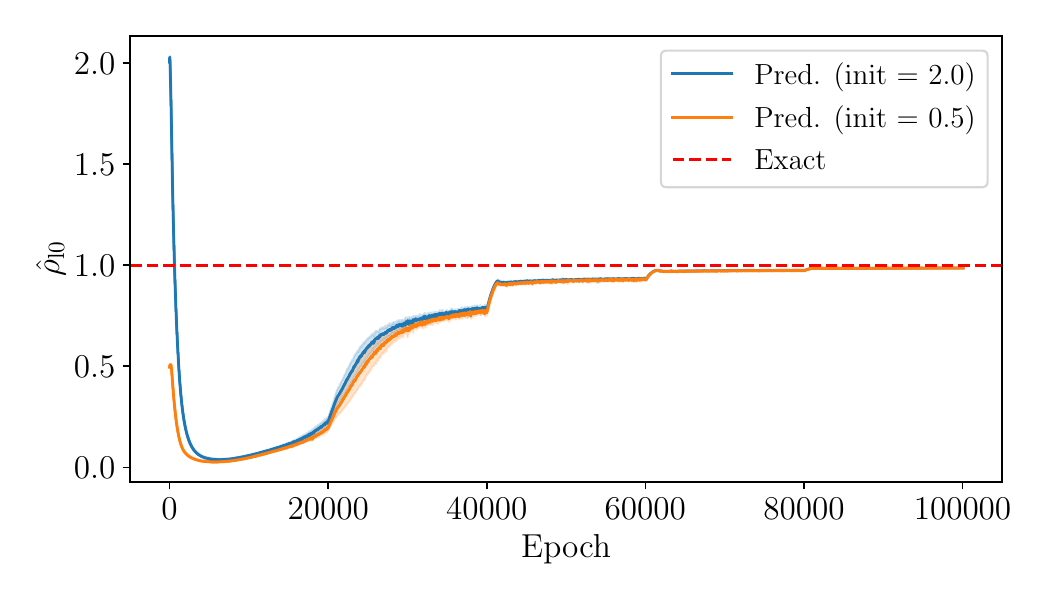}
  \end{minipage}
  \begin{minipage}[t]{0.48\columnwidth}
    \includegraphics[width=\columnwidth]{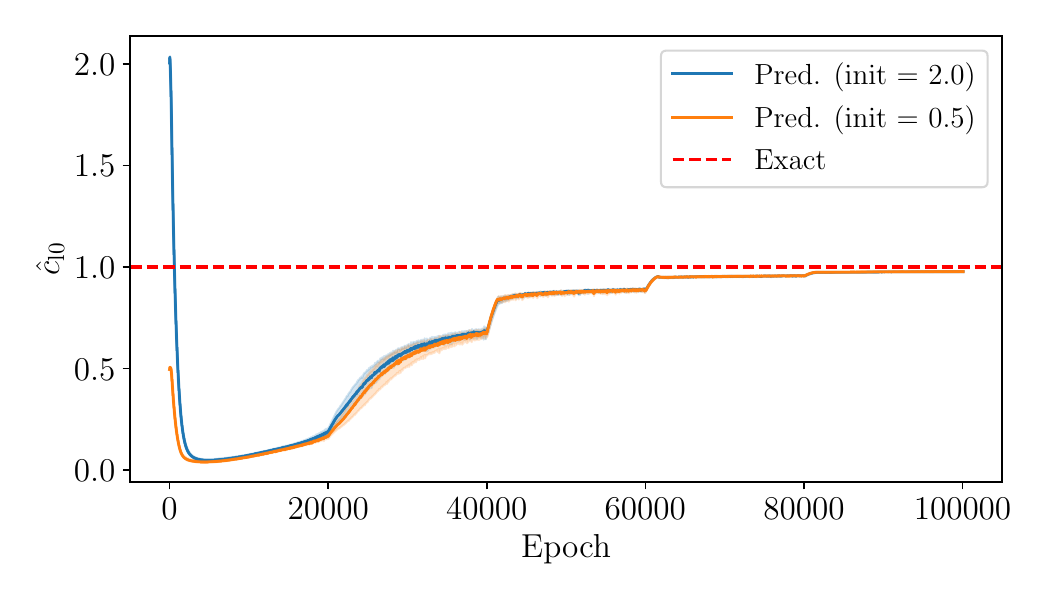}
  \end{minipage}
  \\
  \begin{minipage}[t]{0.48\columnwidth}
    \includegraphics[width=\columnwidth]{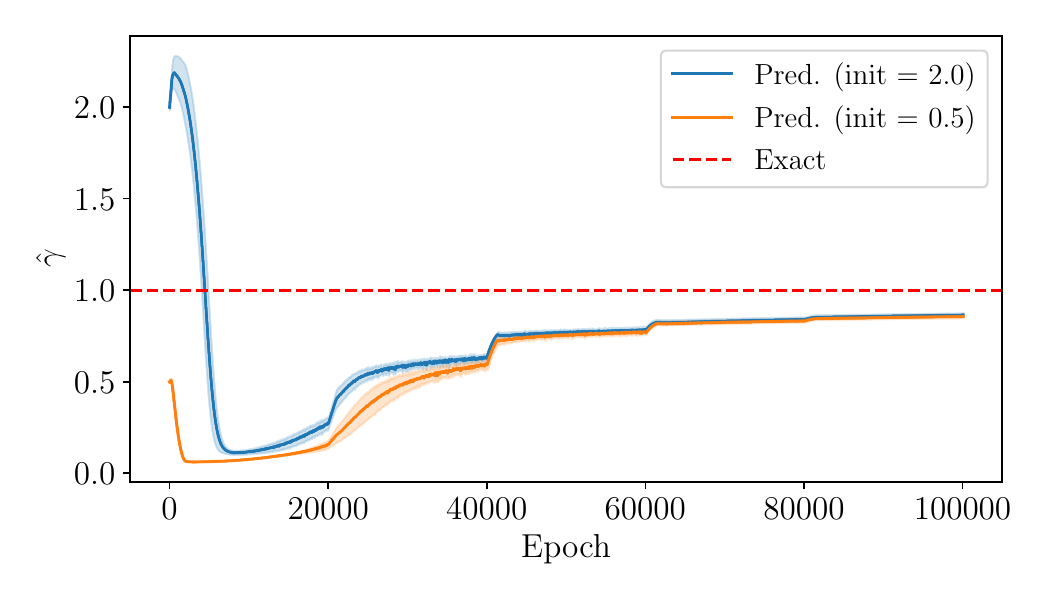}
  \end{minipage}
  \begin{minipage}[t]{0.48\columnwidth}
    \includegraphics[width=\columnwidth]{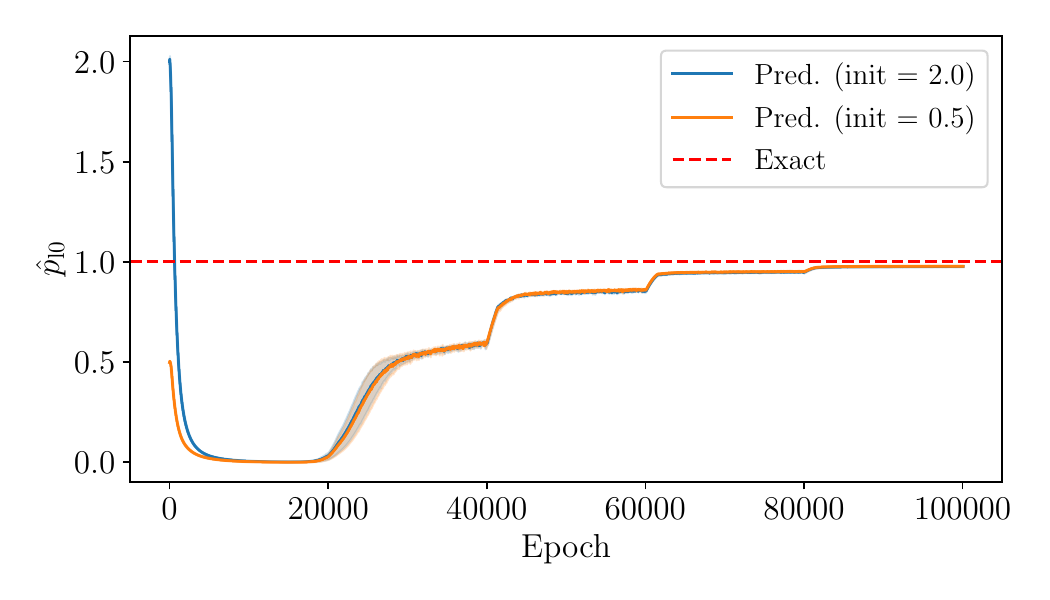}
  \end{minipage}
  \\
  \begin{minipage}[t]{0.48\columnwidth}
    \includegraphics[width=\columnwidth]{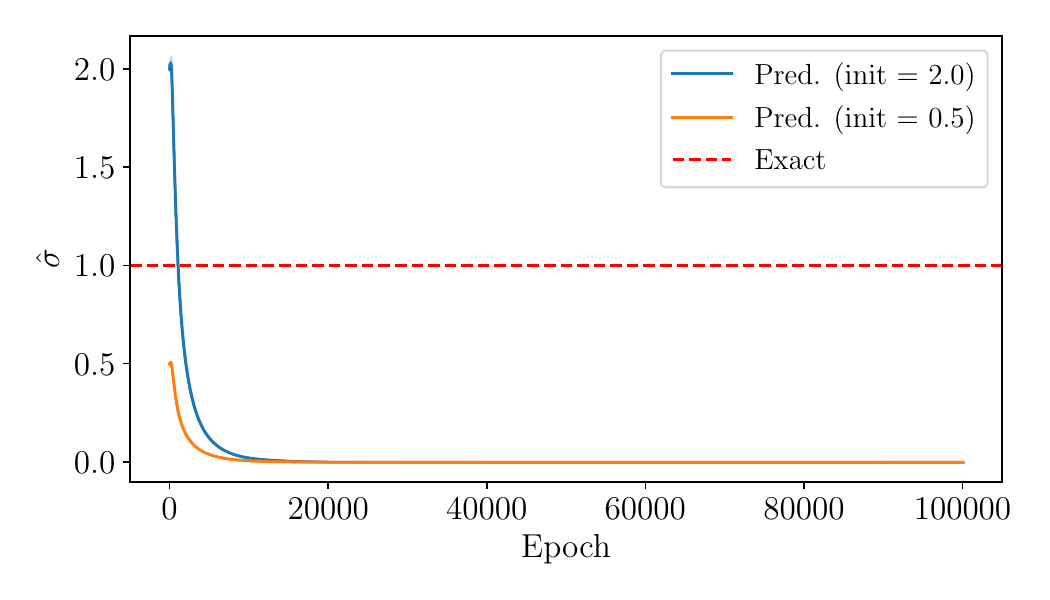}
  \end{minipage}
    \caption{Estimation process of the normalized parameters. Each solid line (blue and orange) represents the mean value over five runs with different random seeds, and the shaded region represents the corresponding standard deviation.}\label{fig:parameter_estimation}
\end{figure}

Figure \ref{fig:parameter_estimation} shows the evolution of the estimated values for the seven parameters. The solid line represents the mean value, and the surrounding shaded region represents the standard deviation; blue and orange correspond to the results obtained with initial values of 2.0 and 0.5, respectively. For every parameter, the converged value was the same regardless of whether the initial value was greater than or less than 1.0. This indicates that, for initial values of approximately 0.5 or 2.0 times the true value, the estimation results follow a similar optimization path irrespective of the initial value. Regarding estimation accuracy, the parameters in Fig.~\ref{fig:parameter_estimation} are arranged in order of the magnitude of the solution sensitivity shown in Fig.~\ref{fig:sensitivity_solution_bar}. For the viscosity of liquid \(\mu_\mathrm{l}\) and the bubble initial radius \(R_0\), which exhibited the largest solution sensitivity, the estimated values converge to approximately 1.0 by 60,000 epochs. For the density of liquid without disturbance \(\rho_\mathrm{l0}\), the sound speed in pure water \(c_\mathrm{l0}\), and the pressure of liquid without disturbance \(p_\mathrm{l0}\), the estimated values also converge to 1.0, but require a larger number of epochs to do so. The polytropic exponent \(\gamma\) approaches 1.0 but does not converge sufficiently close to it. For the surface tension coefficient \(\sigma\), which exhibited the lowest solution sensitivity among the seven parameters, the estimated value converges toward 0 from an early stage of training (in practice, owing to the exponential parameterization enforcing positivity, \(\hat\sigma\) asymptotically approaches, but by construction never reaches or crosses, zero), indicating that this parameter could not be estimated at all. This suggests that, in terms of optimizing the loss function, deviations of \(\sigma\) from its true value have almost no effect on the loss function, and it is precisely this insensitivity that makes \(\sigma\) difficult to estimate.

Table \ref{tbl:parameter_estimation_summary} shows the final estimated parameter values and the corresponding errors after 100,000 epochs of training. For parameters other than \(\gamma\) and \(\sigma\), the estimation accuracy was found to be within approximately 2\%. The entries are arranged in descending order of solution sensitivity; with the exception of \(\gamma\), the error values are found to be arranged in an approximately ascending order as well, indicating that the estimation accuracy is generally consistent with the results of the solution sensitivity analysis.

\begin{table}[t]
\caption{Final estimated values and relative errors of the seven physical parameters after 100,000 training epochs, for initial values of 2.0 and 0.5 times the true value. Values are reported as the mean \(\pm\) standard deviation over five runs with different random seeds.}
\label{tbl:parameter_estimation_summary}
\begin{tabular*}{\tblwidth}{@{}LLLL@{}}
\toprule
 Parameter & Initial value & Prediction (mean \(\pm\) std) & Error (\(\%\)) \\ 
\midrule
 \(\hat{\mu}_\mathrm{l}\)     & 2.0 & 1.0086 \(\pm\) 0.0009 & 0.8599 \\
                              & 0.5 & 1.0085 \(\pm\) 0.0004 & 0.8502 \\ \cmidrule{1-4}
 \(\hat{R}_0\)                & 2.0 & 0.9916 \(\pm\) 0.0004 & 0.8439 \\
                              & 0.5 & 0.9913 \(\pm\) 0.0007 & 0.8652 \\ \cmidrule{1-4}
 \(\hat{\rho}_\mathrm{l0}\)   & 2.0 & 0.9882 \(\pm\) 0.0008 & 1.1842 \\
                              & 0.5 & 0.9880 \(\pm\) 0.0007 & 1.1981 \\ \cmidrule{1-4}
 \(\hat{c}_\mathrm{l0}\)      & 2.0 & 0.9790 \(\pm\) 0.0011 & 2.0963 \\
                              & 0.5 & 0.9794 \(\pm\) 0.0012 & 2.0578 \\ \cmidrule{1-4}
 \(\hat{\gamma}\)             & 2.0 & 0.8668 \(\pm\) 0.0079 & 13.3226 \\
                              & 0.5 & 0.8583 \(\pm\) 0.0110 & 14.1703 \\ \cmidrule{1-4}
 \(\hat{p}_\mathrm{l0}\)      & 2.0 & 0.9793 \(\pm\) 0.0022 & 2.0725 \\
                              & 0.5 & 0.9806 \(\pm\) 0.0023 & 1.9375 \\ \cmidrule{1-4}
 \(\hat{\sigma}\)             & 2.0 & 0.0005 \(\pm\) 0.0000 & 99.9539 \\
                              & 0.5 & 0.0003 \(\pm\) 0.0000 & 99.9737 \\
\bottomrule
\end{tabular*}
\end{table}

\section{Summary} \label{sec:summary}
In this study, we constructed a parameter estimation framework that combines the KdVB equation, which describes nonlinear ultrasound propagation in a bubbly liquid, with a PINN. The contributions of this study are threefold. First, we showed that the PINN can accurately predict the solution of the governing equation in the forward problem. Second, prior to performing inverse estimation, we presented a sensitivity analysis framework that evaluates the sensitivity of both the governing-equation coefficients and the solution itself to each physical parameter, thereby enabling the expected difficulty of estimating each parameter to be quantitatively assessed in advance. Third, building on this prior assessment, we performed single-parameter inverse estimation and confirmed that the resulting estimation accuracy was closely correlated with the sensitivity predicted in advance: parameters with high solution sensitivity were estimated accurately, whereas the parameter with the lowest sensitivity could not be estimated at all.

Across the three numerical experiments, the results were mutually consistent. The forward problem confirmed that the PINN reproduces the reference solution, with a near-zero error at the initial time confirming that the initial condition is enforced exactly as intended. The sensitivity analysis, evaluated at both the coefficient and solution levels, identified the viscosity of liquid \(\mu_\mathrm{l}\) and the bubble initial radius \(R_0\) as the most estimable parameters and the surface tension coefficient \(\sigma\) as the least; for the bubble number density \(N_\mathrm{g}\), we further showed analytically (\ref{sec:independent_of_ng}) that none of the coefficients of the governing equation depend on it, so that its estimation is fundamentally impossible in principle. The inverse problem confirmed this ordering: \(\mu_\mathrm{l}\), \(R_0\), \(\rho_\mathrm{l0}\), \(c_\mathrm{l0}\), and \(p_\mathrm{l0}\) were estimated accurately, with errors of approximately 0.8--2.1\%, whereas \(\gamma\) converged only partially and \(\sigma\) could not be estimated at all. This correlation indicates that the proposed sensitivity analysis can serve not only to anticipate, prior to inverse estimation, which parameters are estimable, but also, more generally, as a diagnostic tool for identifying the likely cause when PINN-based parameter inversion fails to converge. Because the sensitivity analysis framework itself does not depend on any property specific to the KdVB equation, such a diagnostic use could, in principle, be extended to PINN-based inverse problems governed by physical models other than the nonlinear wave equation considered in this study.

In the present study, only the effectiveness of single-parameter estimation was confirmed; for more general applicability, simultaneous estimation of multiple parameters, as well as estimation from limited and noisy observational data, will be required. In particular, since the eight physical parameters enter the governing equation only through a small number of dimensionless parameter groups (\(\Pi_1\), \(\Pi_2\), \(\Pi_3\)), different combinations of parameters may yield similar solutions, which could make it more difficult to separately estimate individual parameters from observational data alone. Investigating this issue, together with application to experimental data, remains a subject for future work.








\section*{Acknowledgement}
This work was supported by JSPS KAKENHI Grant Numbers JP25KJ0658 and JP25KJ0026, and JST FOREST Grant Number JPMJFR215S.

\section*{Declaration of generative AI and AI-assisted technologies in the manuscript preparation process}
During the preparation of this work the authors used Claude (Anthropic) in order to improve the English language and readability of the manuscript. After using this tool, the authors reviewed and edited the content as needed and take full responsibility for the content of the published article.

\appendix
\renewcommand{\thesection}{Appendix \Alph{section}} 
\counterwithin{equation}{section}
\renewcommand{\theequation}{\Alph{section}.\arabic{equation}}

\section{Explicit form of the coefficients \(\Pi_1\), \(\Pi_2\), and \(\Pi_3\) and the nondimensional phase velocity \(v_\mathrm{p}\) \citep{yu2025propagation}} \label{sec:explicit_form}

The nonlinear coefficient \(\Pi_1\), dissipative coefficient \(\Pi_2\), and dispersive coefficient \(\Pi_3\) are given as
\begin{align}
&\Pi_1 
= 
-\frac{1}{6}k_1 + \frac{1}{6\beta_0}k_2 + \frac{1}{6\beta_0(1-\beta_0)}k_3 + k_4, \label{eq:pi1} \\
&\Pi_2
=
\frac{1}{2}\sqrt{\epsilon}v_\mathrm{p}
\left(
\frac{4\omega\mu^\ast_\mathrm{l}}{\epsilon\rho^\ast_\mathrm{l0}R^\ast_0\omega^\ast_\mathrm{B}} + \frac{\omega\omega^\ast_\mathrm{B}R^\ast_0}{\sqrt{\epsilon}c^\ast_\mathrm{l0}}
\right), \label{eq:pi2} \\
&\Pi_3
=
\frac{1}{2}\omega^2 v^2_\mathrm{p}, \label{eq:pi3} \\
&k_1
=
-6\left(b_1-2\right)+2\left(3b_2-b_1b_2\right), \label{eq:k1} \\
&k_2 = -2\beta_0b_1b_2, \label{eq:k2} \\
&k_3 = -2\beta_0b_1b_2-2\left(1-\beta_0\right)b^2_2, \label{eq:k3} \\
&k_4 = 1 + \frac{3\gamma\left(3\gamma-1\right)p^\ast_\mathrm{g0}}{2{R^\ast_0}^2{\omega^\ast_\mathrm{B}}^2\rho^\ast_\mathrm{l0}}, \label{eq:k4} \\
&b_1 = 3\left(1-\beta_0\right), \label{eq:b1} \\
&b_2 = -3\beta_0, \label{eq:b2}
\end{align}
where \(\beta_0 = 4\pi N^\ast_\mathrm{g} {R^\ast_0}^3 / 3\) is the void fraction, \(\omega\) is the nondimensional parameter related to wave frequency, \(p^\ast_\mathrm{g0} = p^\ast_\mathrm{l0} + 2\sigma^\ast/R^\ast_0\) is the pressure of the gas inside a bubble without disturbance and \(\omega^\ast_\mathrm{B}\) is the natural angular frequency of bubbles:
\begin{align}
\omega^\ast_\mathrm{B} = \sqrt{\frac{3\gamma p^\ast_\mathrm{l0}+\left(3\gamma-1\right)\frac{2\sigma^\ast}{R^\ast_0}}{\rho^\ast_\mathrm{l0}{R^\ast_0}^2}}.
\label{eq:bubble_natural_angular_freq}
\end{align}

The nondimensional phase velocity \(v_\mathrm{p}\) is given as
\begin{align}
v_\mathrm{p} = \sqrt{\frac{{R^\ast_0}^2{\omega^\ast_\mathrm{B}}^2}{3\beta_0\left(1-\beta_0\right)\epsilon{U^\ast}^2}},
\label{eq:nondim_phase_velocity}
\end{align}
where \(U^\ast\) is the typical phase velocity of the wave:
\begin{align}
U^\ast = \sqrt{\frac{p^\ast_\mathrm{l0}}{\beta_0\left(1-\beta_0\right)\rho^\ast_\mathrm{l0}}}.
\label{eq:typical_phase_velocity}
\end{align}

\section{Independence of the governing equation on the bubble number density}
\label{sec:independent_of_ng}

We show, using the explicit expressions given in \ref{sec:explicit_form}, that the nonlinear coefficient \(\Pi_1\), the dissipative coefficient \(\Pi_2\), and the dispersive coefficient \(\Pi_3\) are all independent of the bubble number density \(N^\ast_\mathrm{g}\). Since \(N^\ast_\mathrm{g}\) enters Eqs.~(\ref{eq:pi1})--(\ref{eq:typical_phase_velocity}) solely through the void fraction \(\beta_0 = 4\pi N^\ast_\mathrm{g} {R_0^\ast}^3/3\), it suffices to show that \(\Pi_1\), \(\Pi_2\), and \(\Pi_3\) are independent of \(\beta_0\): by the chain rule,
\begin{equation}
\frac{\partial \Pi_i}{\partial N^\ast_\mathrm{g}} = \frac{\partial \Pi_i}{\partial \beta_0}\cdot\frac{\partial \beta_0}{\partial N^\ast_\mathrm{g}}, \qquad \frac{\partial \beta_0}{\partial N^\ast_\mathrm{g}} = \frac{4}{3}\pi {R^\ast_0}^3 \neq 0,
\end{equation}
so \(\partial \Pi_i/\partial \beta_0 = 0\) implies \(\partial \Pi_i/\partial N^\ast_\mathrm{g} = 0\) for \(i=1,2,3\).

Substituting Eqs.~(\ref{eq:b1}) and (\ref{eq:b2}) into Eqs.~(\ref{eq:k1})--(\ref{eq:k3}) and expanding gives
\begin{equation}
-\frac{k_1}{6} = 1-3\beta_0+3\beta_0^2, \qquad
\frac{k_2}{6\beta_0} = 3\beta_0(1-\beta_0) = 3\beta_0-3\beta_0^2, \qquad
k_3 \equiv 0.
\end{equation}
The vanishing of \(k_3\) follows because its two terms are identical up to sign, \(-2\beta_0 b_1b_2 = 18\beta_0^2(1-\beta_0) = 2(1-\beta_0)b_2^2\). Substituting these into Eq.~(\ref{eq:pi1}), the \(\beta_0\)- and \(\beta_0^2\)-terms cancel exactly, leaving
\begin{equation}
\Pi_1 = 1 + k_4.
\end{equation}
Since \(k_4\) (Eq.~(\ref{eq:k4})) is a function of \(\gamma\), \(p^\ast_\mathrm{g0}\), \(R^\ast_0\), \(\omega^\ast_\mathrm{B}\), and \(\rho^\ast_\mathrm{l0}\) alone, none of which depend on \(\beta_0\), \(\Pi_1\) is independent of \(\beta_0\), and hence of \(N^\ast_\mathrm{g}\).

From Eqs.~(\ref{eq:nondim_phase_velocity}) and (\ref{eq:typical_phase_velocity}),
\begin{equation}
v^2_\mathrm{p} = \frac{{R_0^\ast}^2{\omega^\ast_\mathrm{B}}^2}{3\beta_0(1-\beta_0)\epsilon {U^\ast}^2}, \qquad {U^\ast}^2 = \frac{p^\ast_\mathrm{l0}}{\beta_0(1-\beta_0)\rho^\ast_\mathrm{l0}}.
\end{equation}
Substituting \({U^\ast}^2\) into \(v^2_\mathrm{p}\), the factor \(\beta_0(1-\beta_0)\) cancels exactly between numerator and denominator:
\begin{equation}
v^2_\mathrm{p} = \frac{{R^\ast_0}^2{\omega^\ast_\mathrm{B}}^2\rho^\ast_\mathrm{l0}}{3\epsilon p^\ast_\mathrm{l0}}.
\end{equation}
Thus \(v_\mathrm{p}\) itself is independent of \(\beta_0\). Since \(\Pi_3 = \tfrac{1}{2}\omega^2 v^2_\mathrm{p}\) (Eq.~(\ref{eq:pi3})), \(\Pi_3\) is independent of \(\beta_0\), and hence of \(N^\ast_\mathrm{g}\).

In Eq.~(\ref{eq:pi2}),
\begin{equation}
\Pi_2 = \frac{1}{2}\sqrt{\epsilon}v_\mathrm{p}\left(\frac{4\omega\mu^\ast_\mathrm{l}}{\epsilon\rho^\ast_\mathrm{l0}{R^\ast_0}\omega^\ast_\mathrm{B}}+\frac{\omega\omega^\ast_\mathrm{B}R^\ast_0}{\sqrt{\epsilon}c^\ast_\mathrm{l0}}\right),
\end{equation}
the bracketed term contains only \(\mu^\ast_\mathrm{l}\), \(\rho^\ast_\mathrm{l0}\), \(R^\ast_0\), \(\omega^\ast_\mathrm{B}\), \(c^\ast_\mathrm{l0}\), \(\omega\), and \(\epsilon\), none of which depend on \(\beta_0\). The only possible dependence on \(\beta_0\) is through \(v_\mathrm{p}\) in the prefactor, which was shown above to be independent of \(\beta_0\). Hence \(\Pi_2\) is independent of \(\beta_0\), and hence of \(N^\ast_\mathrm{g}\).

Since \(\Pi_1\), \(\Pi_2\), and \(\Pi_3\) are all independent of \(\beta_0\), and \(N^\ast_\mathrm{g}\) enters these coefficients solely through \(\beta_0\), none of the coefficients in the governing equation~(\ref{eq:kdvb}) depend on \(N^\ast_\mathrm{g}\). Moreover, the initial condition \(u_0(x)=0.1\sin(\pi x)\) and the periodic boundary condition in Eq.~(\ref{eq:kdvb}) contain no physical parameters at all. The governing equation, together with the initial and boundary conditions, is therefore entirely independent of \(N^\ast_\mathrm{g}\), so the solution \(u(x,t)\) cannot depend on \(N^\ast_\mathrm{g}\) either. Consequently, estimating \(N^\ast_\mathrm{g}\) from observations of \(u(x,t)\) is fundamentally impossible in principle, consistent with the null sensitivity of \(N_g\) observed in both the coefficients (Fig.~\ref{fig:sensitivity_heatmap}) and the solution (Fig.~\ref{fig:sensitivity_solution_bar}).

\bibliographystyle{cas-model2-names}

\bibliography{references}



\end{document}